\documentclass[reprint,10pt,aps,prb,floatfix]{revtex4-1}
\usepackage{amsmath}    
\usepackage{graphicx}   

\begin{document}

\title{Trajectories of charged particles trapped in Earth's magnetic field}
\author{M. Kaan \"Ozt\"urk}
\affiliation{Yeditepe University, Information Systems and Technologies, Istanbul, Turkey}
\email{kaan.ozturk@yeditepe.edu.tr, mkozturk@yahoo.com}
\date{\today}

\begin{abstract}
I outline the theory of relativistic charged-particle motion in the magnetosphere in a way suitable for undergraduate courses. I discuss particle and guiding center motion, derive the three adiabatic invariants associated with them, and present particle trajectories in a dipolar field. I provide twelve computational exercises that can be used as classroom assignments or for self-study. Two of the exercises, drift-shell bifurcation and Speiser orbits, are adapted from active magnetospheric research. The Python code provided in the supplement can be used to replicate the trajectories and can be easily extended for different field geometries.
\end{abstract}

\maketitle

\section{Introduction}
Ions and electrons trapped in the Earth's magnetic field may affect our technology and our daily lives in significant ways. Energetic plasma particles may penetrate satellites and disable them temporarily or permanently. They can also pose serious health hazards for astronauts in space. Spectacles like the aurora are created by particles that enter the Earth's atmosphere at polar regions; on the other hand, aircraft personnel and frequent flyers may accumulate a significant dose of radiation due to the same particles.\cite{Townsend2001} All of these effects are enhanced at periods of solar maximum, the next one being expected to happen between 2012 and 2014. Occasional extreme solar events may induce currents in the ionosphere, which in turn induce significant currents on power lines, causing power outages.\cite{Moldwin2008} Such events can also disrupt communications, radio and GPS. Thus, understanding and predicting the processes in the Earth's magnetosphere have practical importance.

This paper aims to outline one of these processes, charged-particle motion and associated adiabatic invariants, for physics students and instructors who wish to use it in lectures. The emphasis is on numerical computation and visualization of trajectories. For a more comprehensive discussion advanced texts on plasma physics\cite{GurBat2005,Sturrock1994,Walt1994,Roederer1970} can be consulted.

Other authors have also suggested using topics from plasma research to enhance undergraduate curriculum. Lopez\cite{Lopez2008} provides examples of how space physics can be incorporated in undergraduate electromagnetism courses, and McGuire\cite{McGuire2003} shows how computer algebra systems can be used to follow particle trajectories in electric dipole and (separately) in magnetic dipole fields. Photographs of plasma experiments, such as those provided by Huggins and Lelek\cite{HugLelek1979} and by the UCLA Plasma Lab web site\cite{plasmalabwww} are also helpful for understanding space plasma behavior. 

\begin{figure}
\includegraphics[scale=0.5]{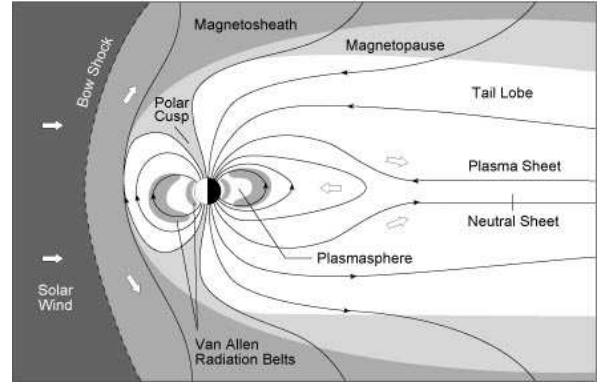}
\caption{\label{fig:magnetosphere}A schematic view of the Earth's magnetosphere. The solar wind comes from the left. (Courtesy of Kenneth R. Lang,\cite{Lang2011} reproduced with permission.)}
\end{figure}

Figure~\ref{fig:magnetosphere} shows a schematic description of the Earth's magnetosphere, which is the region in space where the magnetic field of the Earth is dominant. Charged particles trapped in the magnetosphere form the radiation belts, the plasmasphere, and current systems such as the ring current, tail current, and field-aligned currents.

The Earth radius~$R_{\rm e}$ (6378.137 km) is a natural length scale for the magnetosphere. Near the Earth, up to 3-4$R_{\rm e}$, the field can be very well approximated with the field of a dipole. However, at larger distances, the effects of the solar wind cause significant deviations from the dipole.

The solar wind is a stream of plasma carrying magnetic field from the Sun. When the solar wind encounters the Earth's magnetosphere, the two systems do not mix. This is because of the ``frozen-in flux'' condition\cite{Kivelson1995} which dictates that plasma particles stay attached to magnetic field lines, except at special locations such as polar cusps. The solar wind influences the magnetosphere by applying mechanical and magnetic pressure on it, compressing it earthward on the side facing the Sun (the ``dayside''). This compression is stronger when the Sun is more active. On the opposite side (the ``nightside''), the field is extended over a very large distance, forming the magnetotail. Wolf\cite{Wolf1995} provides a review of the complex and time-dependent interactions between magnetic fields, induced electric fields and plasma populations.

The Van Allen radiation belts form a particularly significant plasma population due to their high energy and their proximity to Earth. They can be found from 1000km above the ground up to a distance of 6$R_{\rm e}$. These belts are composed of electrons with energies up to several MeVs and of protons with energies up to several hundred MeVs. The dynamics of these particles is the main focus of this paper.

This paper is organized as follows: Section~\ref{sec:NLeqs} introduces the relativistic equation of motion for a particle in an electric and magnetic field and describes the cyclotron, bounce and drift motions. It also shows some typical particle trajectories under the dipolar magnetic field, approximating the Earth's field. Section~\ref{sec:firstinv} introduces the concept of adiabatic invariants and derives the first adiabatic invariant associated with the particle motion. Section~\ref{sec:gceqs} gives the approximate equations of motion for the guiding center of a particle, obtained by averaging out the cyclotron motion. Section~\ref{sec:secondthirdinv} presents and derives the second and third invariants associated with the bounce and drift motions, respectively. Section~\ref{sec:exercises} lists some exercises building on the concepts described in the paper. Two of these exercises describe non-dipole fields that are used for modeling different regions of the magnetosphere.

\section{Particle trajectory in dipolar magnetic field}
\label{sec:NLeqs}
The motion of a particle with charge $q$ and mass $m$ in an electric field $\mathbf{E}$ and magnetic field $\mathbf{B}$ is described by the Newton-Lorentz equation:
\begin{equation}
\label{eq:NLeqs}
\frac{\mathrm{d}(\gamma m \mathbf{v})}{\mathrm{d}t} = q\mathbf{E}(\mathbf{r}) + q\mathbf{v}\times\mathbf{B}(\mathbf{r}).
\end{equation}
Here $\gamma = (1-v^2/c^2)^{-(1/2)}$ is the relativistic factor and $v$ is the particle speed. 

Suppose that $\mathbf{E}=\mathbf{0}$. Then, because of the cross product, the acceleration of the particle is perpendicular to the velocity at all times, so the speed of the particle (and the factor $\gamma$) remains constant. Further suppose that the magnetic field is uniform. Then, particles move on helices parallel to the field vector. The circular part of this motion is called the ``cyclotron motion'' or the ``gyromotion''. The ``cyclotron frequency'' $\Omega$ and the ``cyclotron radius'' $\rho$ are respectively given by
\begin{eqnarray}
\label{eq:cycfreq}\Omega &=& \frac{qB}{\gamma m}\\
\label{eq:cycrad}\rho &=& \frac{\gamma mv_\perp}{qB},
\end{eqnarray}
where $B=|\mathbf{B}|$ is the uniform field strength and $v_\perp$ is the component of the velocity perpendicular to the field vector. If there are not any other forces, the parallel component of the velocity remains constant. The combined motion traces a helix.\cite{Griffiths1999} 

If the electric field is not zero, we can write it as $\mathbf{E}=\mathbf{E}_\perp + \mathbf{E}_{||}$, where $\mathbf{E}_{||}$ is parallel to $\mathbf{B}$, and $\mathbf{E}_\perp$ is perpendicular to it. If $\mathbf{E}_{||}\neq\mathbf{0}$, particles accelerate with $q\mathbf{E}_{||}/m$ along the field line and they are rapidly removed from the region. Therefore, the existence of a trapped plasma population implies that the parallel electric field must be negligible.

The perpendicular component of the electric field will move particles with an overall drift velocity, known as the E-cross-B-drift, which is perpendicular to both field vectors:
\begin{equation}
 \mathbf{v}_E = \frac{\mathbf{E}_\perp\times\mathbf{B}}{B^2}.
\end{equation}
The particle will move with with the velocity $\mathbf{v}_E$, plus the cyclotron motion described above. The drift velocity $\mathbf{v}_E$ is independent of particle mass and charge. Therefore, in an inertial frame moving with $\mathbf{v}_E$, the E-cross-B-drift will vanish for all types of particles.

For the remainder of this paper we take $\mathbf{E}=0$, that is, the acceleration due to the electric field is not taken into consideration. This is not because electric fields are unimportant; on the contrary, they play an important role in the complex dynamics of plasmas. The first reason for leaving out electric field effects is described above: If the field is uniform and constant, we can transform to another frame that cancels it. Even if the field is nonuniform and time-dependent, electric drifts can be vectorially added to magnetic drifts in order to obtain the overall drift. Drift velocities due to different fields are independent.

The second reason is the need for simplicity; a static magnetic field provides sufficient real-life context for the discussion of guiding-center and adiabatic invariant concepts in general-purpose lectures. The final reason is that this paper focuses on the region occupied by radiation belts, and in this region the magnetic term of~(\ref{eq:NLeqs}) is the dominant force.\cite{note:neglectofE}

Now we consider motion under the influence of a magnetic dipole. The field $\mathbf{B}_{\rm dip}(\mathbf{r})$ of a magnetic dipole with moment vector $\mathbf{M}$ at location $\mathbf{r}$ is given by:\cite{Griffiths1999}
\begin{equation}
 \mathbf{B}_\mathrm{dip}(\mathbf{r}) = \frac{\mu_0}{4\pi r^3} \left[ 3(\mathbf{M}\cdot\hat{\mathbf{r}})  \hat{\mathbf{r}} - \mathbf{M} \right],
\end{equation}
where $\mathbf{r} = x\hat{\mathbf{x}}+y\hat{\mathbf{y}}+z\hat{\mathbf{z}}$, $r=|\mathbf{r}|$ and $\hat{\mathbf{r}}=\mathbf{r}/r$. For Earth, we take $\mathbf{M} = -M\hat{\mathbf{z}}$, antiparallel to the $z$-axis, because the magnetic north pole is near the geographic south pole.\cite{note:GSE}

At the magnetic equator ($x=1R_{\rm e}$, $y=z=0$) the field strength is measured to be $B_0 = 3.07\times10^{-5}\mathrm{T}$. Substitution shows that $\mu_0 M/4\pi = B_0R_{\rm e}^3$. Then in Cartesian coordinates, the field is given by:
\begin{equation}
\label{eq:dipolefield}
 \mathbf{B}_\mathrm{dip} = -\frac{B_0R_{\rm e}^3}{r^5} \left[ 3xz\hat{\mathbf{x}} + 3yz\hat{\mathbf{y}} + (2z^2 - x^2 -y^2)\hat{\mathbf{z}} \right].
\end{equation}

\begin{figure}
 \includegraphics{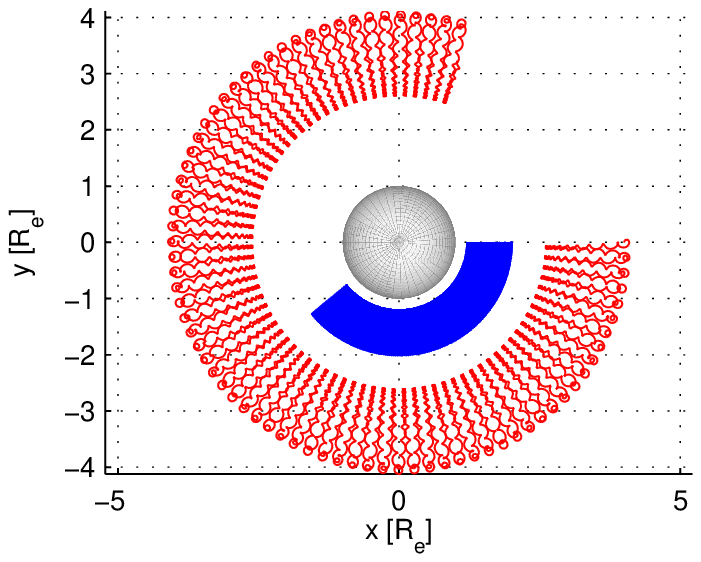}
 \includegraphics{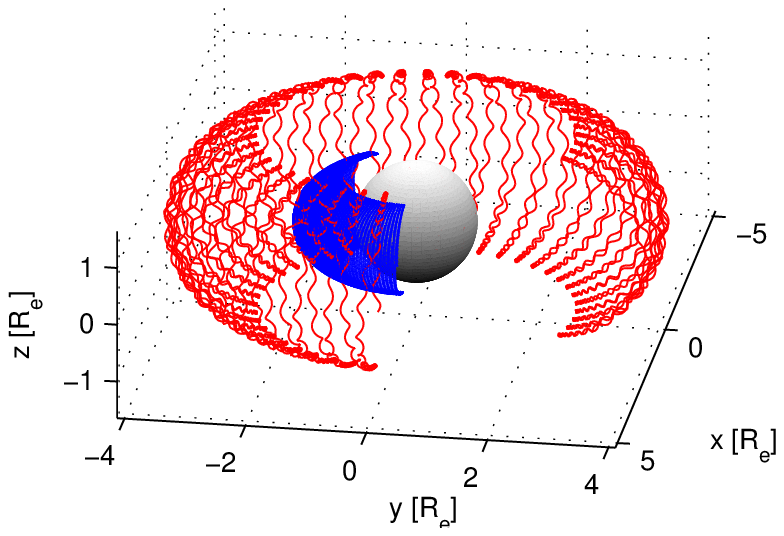}
 \caption{\label{fig:proton_full}(Color online) Trajectories of two 10MeV protons in the Earth's dipole field. The dipole moment is in the $-\hat{\bf z}$ direction. Both panels show the same trajectories from different viewing angles.}
\end{figure}

Figure~\ref{fig:proton_full} shows trajectories of two protons with 10MeV kinetic energy, a typical energy for radiation belts.\cite{Wolf1995} The trajectories are calculated with the SciPy module using the Python language.\cite{supp} One proton is started at $(2R_\mathrm{e}, 0,0)$ and the other at $(4R_\mathrm{e}, 0,0)$. Both start with an equatorial pitch angle (angle between the velocity and field vectors) $\alpha_{\rm eq} = 30^\circ$ so that $v_{y0}=v\sin\alpha_{\rm eq}$, $v_{z0}=v\cos\alpha_{\rm eq}$ and $v_{x0}=0$. Both are followed for 120 seconds. 

The motion is again basically helical, but the nonuniformity of the field introduces two additional modes of motion on large spatial and temporal scales. These are called ``the bounce motion'' and ``the drift motion''.

The bounce motion proceeds along the field line that goes through the helix (the ``guiding line''). The motion slows down as it moves toward locations with a stronger magnetic field, reflecting back at ``mirror points''. The bounce motion is much slower than the cyclotron motion.

The drift motion takes the particle across field lines (perpendicular to the bounce motion). In general, drift motion is faster at larger distances, as observed in Figure~\ref{fig:proton_full}. Particles in dipole-like fields are trapped on closed ``drift shells'' as long as they are not disturbed by collisions or interactions with EM waves. The drift motion is much slower than the bounce motion. 

Under a dipolar field, the bounce motion period $\tau_{\rm b}$ and the drift motion period $\tau_{\rm d}$ are approximately given as:\cite{Walt1994}
\begin{eqnarray}
 \label{eq:bounceperiod}\tau_{\rm b} &\approx& 0.117\left(\frac{R_0}{R_{\rm e}}\right) \frac{c}{v} \left[1-0.4635(\sin \alpha_{\rm eq})^{3/4}\right] \\
 \label{eq:driftperiod}\tau_{\rm d} &\approx& \frac{2\pi qB_0R_{\rm e}^3}{mv^2}\frac{1}{R_0}\left[1-\frac{1}{3}(\sin \alpha_{\rm eq})^{0.62}\right],
\end{eqnarray}
where $R_0$ is the equatorial distance to the guiding line and $\alpha_{\rm eq}$ is the equatorial pitch angle. Both approximations have an error of about 0.5\%.

\section{The first adiabatic invariant}
\label{sec:firstinv}
If the parameters of an oscillating system are varied very slowly compared to the frequency of oscillations, the system possesses an ``adiabatic invariant'', a quantity that remains approximately constant. In the Hamiltonian formalism, the adiabatic invariant is the same as the action variable:\cite{HandFinch1998}
\begin{equation}
 J = \oint_{H=E} p(E)\,\mathrm{d}q,
\end{equation}
where $q$, $p$ are canonical variables and the integral is evaluated over one cycle of motion satisfying $H(p,q,t)=E$. The integral should be evaluated at ``frozen time'', that is, the slowly varying parameter is considered constant during the integration cycle.

There are three separate periodic motions of a charged particle in a dipole-like magnetic field. This means there are three adiabatic invariants for the particle's motion.

The canonical momentum for a charged particle in a magnetic field with vector potential $\mathbf{A}$ is $\gamma m\mathbf{v}+q\mathbf{A}$. To obtain the first adiabatic invariant $J_1$, we integrate the canonical momentum over one cycle of the cyclotron orbit:
\begin{equation}
 J_1 = \oint (\gamma m\mathbf{v}+q\mathbf{A})\cdot\mathrm{d}\boldsymbol{\ell}.
\end{equation}
Here $\mathrm{d}\boldsymbol{\ell}$ is the line element of the particle trajectory. Even though the path does not exactly close, we evaluate the integral as if it does. Stokes' Theorem states that: 
\begin{equation}
 \oint \mathbf{A}\cdot\mathrm{d}\boldsymbol{\ell} = \int \nabla\times \mathbf{A}\cdot\mathrm{d}\boldsymbol{\sigma},
\end{equation}
where the right-hand side integral is taken over the surface bounded by the closed path of the left-hand side integral. Using Stokes' theorem with $\mathbf{B} = \nabla\times \mathbf{A}$, the integral $J_1$ takes the form:
\begin{eqnarray}
J_1 &=& 2\pi \gamma m \rho v_\perp - q \pi \rho^2 B \\
 &=& \frac{\pi \gamma^2 m^2 v_\perp^2}{qB}.
\end{eqnarray}
The second equation follows from substituting $\rho$ from~(\ref{eq:cycrad}). It is assumed that the cycle is sufficiently small so that $\mathbf{B}$ is considered uniform over the area bounded by the gyromotion. This assumption is essential for the existence of adiabatic invariants. The area element $ \mathrm{d}\boldsymbol{\sigma}$ is antiparallel to $\mathbf{B}$ because of the sense of gyration of the particles; hence the negative sign of the second term. 

Customarily, one takes 
\begin{equation}
\label{eq:magneticmoment}
 \mu = \frac{\gamma^2 m v_\perp^2}{2B}
\end{equation}
as the first invariant, which differs from $J_1$ only in some constant factors. The parameter $\mu$ is called the ``magnetic moment'' because it is equal to the magnetic moment of the current generated by the particle moving on this circular path.

The adiabatic invariance of $\mu$ explains the existence of mirror points. As the particle moves along the field line toward points with larger $B$, the perpendicular speed $v_\perp$ must increase in order to keep $\mu$ constant. However, $v_\perp$ can be at most $v$, the constant total speed. The particle stops at the point where $B=B_{\rm m}=\gamma^2 m v^2/2\mu$ and falls back (see Section~\ref{sec:gceqs}).

\begin{figure*}
\includegraphics{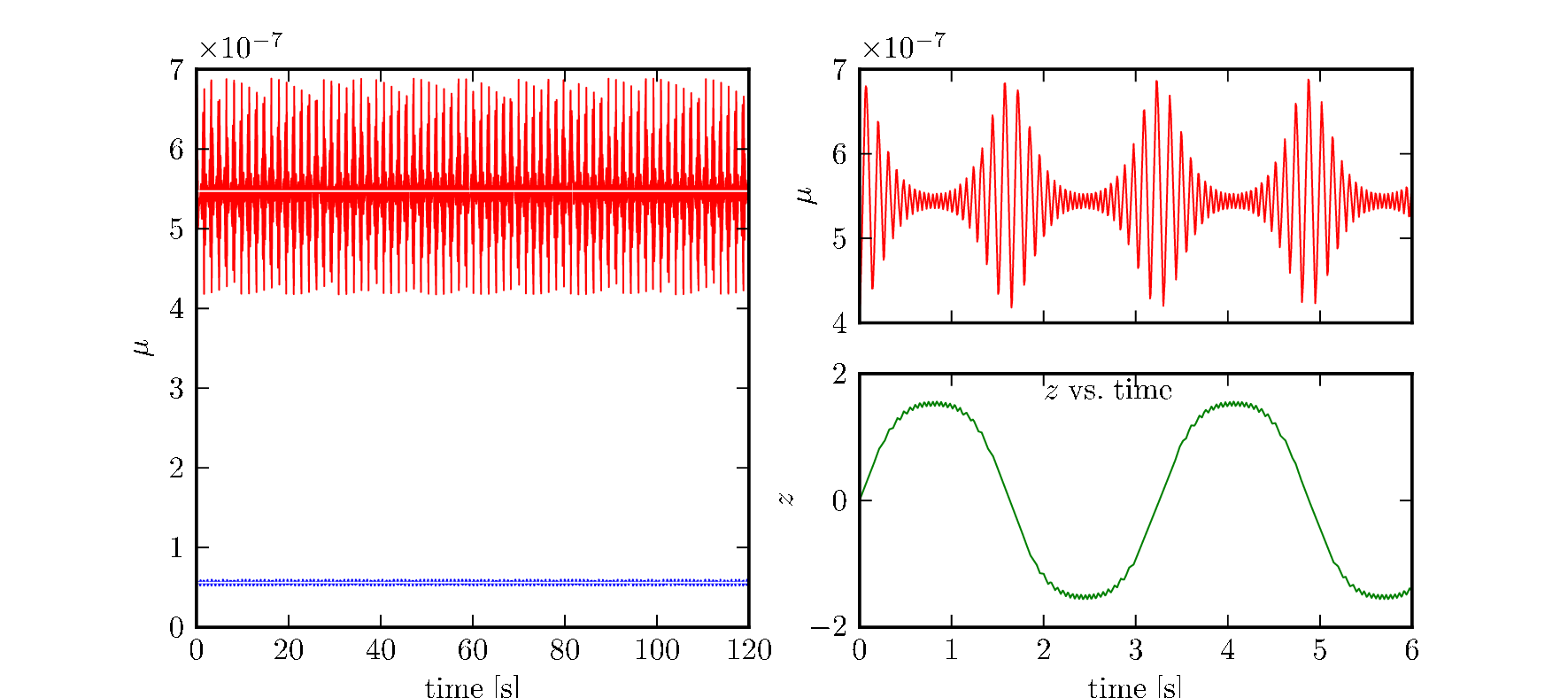}
 \caption{\label{fig:firstinv}(Color online) Left: The instantaneous values of the magnetic moment for the particle orbits shown in Fig. \ref{fig:proton_full}. Lower curve is for the proton starting at 2 $R_{\rm e}$ distance, upper curve for the proton starting at 4 $R_{\rm e}$. The white horizontal line shows the average value. Top right: A close-up view of the instantaneous magnetic moment up to time 6s. Bottom right: The $z$-coordinate up to time 6s.}
\end{figure*}

The left panel of Figure~\ref{fig:firstinv} shows the magnetic moment $\mu$ for the two protons shown in Figure~\ref{fig:proton_full}. The values are oscillating with the local cyclotron frequency because instantaneous values of $v_\perp$ and $B$ are used. The actual adiabatic invariant is the average of these oscillations and it is constant in time.

The top right panel shows the oscillations of $\mu$ vs. time for a shorter interval. Comparison with the $z$ vs. time plot below, it can be seen that these oscillations are correlated with the bounce motion. Oscillations of $\mu$ have a small amplitude near the mirror point because there the parallel motion slows down and the overall motion becomes more adiabatic. Near the equatorial plane ($z=0$) parallel motion is fastest, the motion is less adiabatic, and $\mu$ oscillates with a larger amplitude.

The proper way of calculating the first invariant would remove all oscillations: After following the full trajectory, find the times $\lbrace t_i \rbrace$ where $v_x=0$ (or $v_y=0$) by searching along discrete path points and by interpolating. The difference between any successive time points is half a gyroperiod. Then take the average of $\mu$ over that time interval using the path points. Repeating this procedure for all time intervals, we get a constant set of $\mu$ values (apart from numerical errors).

\section{The guiding-center equations}
\label{sec:gceqs}
The ``guiding center'' is the geometric center of the cyclotron motion. If the magnetic field is uniform the guiding center moves with constant velocity parallel to the field line. In nonuniform field geometries, there is a sideways drift in addition to the motion along the field lines, as seen in the dipole example in Figure~\ref{fig:proton_full}.

Calculation of the guiding center motion requires that the motion is helical in the smallest scale, and that the field does not change significantly within a cyclotron radius. This condition can be expressed as
\begin{equation}
\label{eq:adbcondition}
 \rho \ll \frac{B}{|\boldsymbol{\nabla}B|}
\end{equation}
The magnetic moment is an adiabatic invariant under this condition.

Northrop\cite{Northrop1963} and Walt\cite{Walt1994} give detailed derivations of the equations of guiding-center motion. In order to derive the acceleration of the guiding center, the particle position $\mathbf{r}$ is substituted with:
\begin{equation}
 \mathbf{r}=\mathbf{R} + \boldsymbol{\rho}
\end{equation}
where $\mathbf{R}$ is the position of the guiding center. The vector $\boldsymbol{\rho}$ lies on the plane perpendicular to the field, oscillates with the cyclotron frequency, and its length is equal to the cyclotron radius.

Assuming that the cyclotron radius is much smaller than the length scale of the field, we can expand $\mathbf{B}(\mathbf{r})$ around $\mathbf{R}$ to first order in a Taylor series:
\begin{equation}
 \mathbf{B}(\mathbf{r}) \approx \mathbf{B}(\mathbf{R}) + (\boldsymbol{\rho}\cdot\boldsymbol{\nabla})\mathbf{B}
\end{equation}

This expansion is substituted into the Newton-Lorenz equation~(\ref{eq:NLeqs}) and the equation is averaged over a cycle, eliminating rapidly oscillating terms containing $\boldsymbol{\rho}$ and its derivatives. The resulting acceleration of the guiding center is given by:
\begin{equation}
\label{eq:GCaccel}
\ddot{\mathbf{R}} = \frac{q}{\gamma m} \dot{\mathbf{R}} \times \mathbf{B}(\mathbf{R}) - \frac{q\rho^2\Omega}{2\gamma m} \boldsymbol{\nabla} B(\mathbf{R})
\end{equation}

Taking the dot product of both sides of the equation with $\hat{\mathbf{b}}$, the local magnetic field direction, will yield the equation of motion along the field line. The first term becomes identically zero because it is perpendicular to the field vector. Then:
\begin{equation}
 \frac{\mathrm{d}v_{||}}{\mathrm{d}t} = - \frac{q\rho^2\Omega}{2\gamma m}\, \hat{\mathbf{b}}\cdot\boldsymbol{\nabla} B(\mathbf{R})
\end{equation}
where $v_{||}$ is the speed along the field line. Replacing $q\rho^2\Omega/(2\gamma m) = \mu/(\gamma^2 m)$ and defining $s$ as the distance along the field line, this equation can be written as:
\begin{equation}
 \frac{\mathrm{d}v_{||}}{\mathrm{d}t} =\frac{\mathrm{d}^2 s}{\mathrm{d} t^2} = -\frac{\partial}{\partial s} \frac{\mu B(s)}{\gamma^2 m}
\end{equation}
Here $B(s)$ is the field strength along the field line. The factors $\mu$ and $\gamma$ can be taken inside the derivative because they are constants. This expression shows that the quantity $\mu B(s)/(\gamma^2 m)$ acts like a potential energy in the parallel direction. The negative sign indicates that the parallel motion is accelerated toward regions with smaller field strength.

The motion of the guiding center perpendicular to field lines can be determined by taking the cross product of Eq.~(\ref{eq:GCaccel}) with the field direction vector. The resulting equation is then iterated to obtain an approximate solution for the drift velocity across field lines.
\begin{equation}
\label{eq:driftvelocity}
 \frac{\mathrm{d}\mathbf{R}_\perp}{\mathrm{d}t} = \frac{\gamma m}{2qB^2}(v^2+v_{||}^2)\,\hat{\mathbf{b}}\times\boldsymbol{\nabla}B
\end{equation}
This drift velocity is actually the sum of two separate drift velocities: The gradient drift that arises from the nonuniformity of the magnetic field, and the curvature drift that occurs because the field lines are curved. For an example of pure gradient drift motion, see Exercise~\ref{ex:graddrift} in Section~\ref{sec:exercises}.

Equation~(\ref{eq:driftvelocity}) shows that electrons and ions drift in opposite directions. This creates a net current around the Earth, called ``the ring current''.

Gradient and curvature drifts are the only drifts seen in static magnetic fields. External electric fields, external forces such as gravity, and time dependent fields create additional drift velocities.\cite{GurBat2005}

Combining these, we obtain the following equations of motion for the guiding center:
\begin{eqnarray}
\label{eq:GCeom}
 \frac{\mathrm{d}\mathbf{R}}{\mathrm{d}t} &=& \frac{\gamma m v^2}{2qB^2}\left(1+\frac{v_{||}^2}{v^2}\right) \hat{\mathbf{b}}\times\boldsymbol{\nabla}B + v_{||}\hat{\mathbf{b}} \\
 \frac{\mathrm{d}v_{||}}{\mathrm{d}t} &=& -\frac{\mu}{\gamma^2 m} \hat{\mathbf{b}}\cdot\boldsymbol{\nabla}B
\end{eqnarray}

These equations are more complicated than the simple Newton-Lorentz equation, and they require computing $\hat{\mathbf{b}}\cdot\boldsymbol{\nabla}B$ and $\hat{\mathbf{b}}\times\boldsymbol{\nabla}B$ at each integration step. Still, they have the advantage that we can follow the overall motion with relatively large time steps because we do not need to resolve the cyclotron motion. This reduces the cumulative error, as well as the total computation time.

\begin{figure}
 \includegraphics{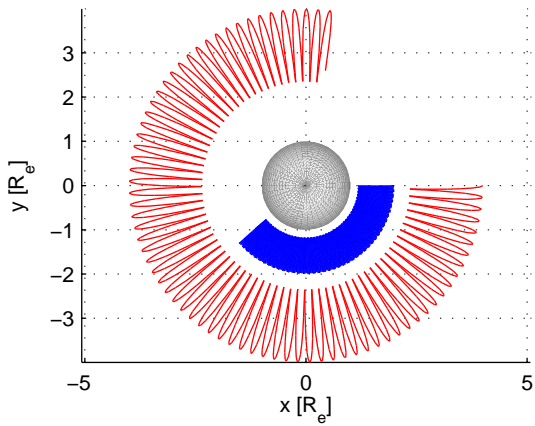}
 \includegraphics{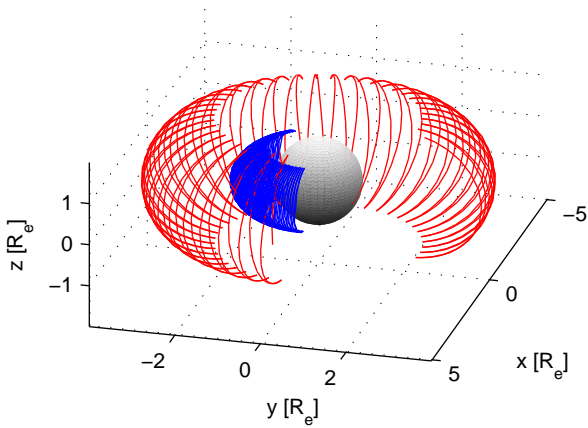}
 \caption{\label{fig:proton_gc}(Color online) Guiding-center trajectories for the particles shown in Figure~\ref{fig:proton_full}. The cyclotron motion is averaged out.}
\end{figure}

Figure~\ref{fig:proton_gc} shows the solution of the guiding-center equations for the same protons shown in Figure~\ref{fig:proton_full} under a dipolar magnetic field (Python source code provided in Supplement).\cite{supp}

It should be noted that the guiding-center equations are approximate because only terms first order in cyclotron radius are used in their derivation. For particles with larger cyclotron radii (higher kinetic energies), there may be a noticeable difference between guiding-center and full-particle trajectories (see Exercise~\ref{ex:gcandfull} in Section~\ref{sec:exercises}). 

\section{The second and third adiabatic invariants}
\label{sec:secondthirdinv}
The second adiabatic invariant is associated with the bounce motion, and it is calculated by integrating the canonical momentum over a path along the guiding field line:
\begin{equation}
 J_2 = \oint (\gamma m \mathbf{v} + q\mathbf{A})\cdot \mathrm{d} \mathbf{s},
\end{equation}
where $\mathrm{d} \mathbf{s}$ is the line element along the field line. The adiabatic integrals are evaluated in a ``frozen'' system: It is assumed that the drift is stopped, so the motion moves back and forth along a single guiding field line.

Using Stokes' theorem, the second term can be converted to an integral over a surface bounded by the bounce path
\begin{equation}
 \oint q\mathbf{A}\cdot \mathrm{d} \mathbf{s} = q\int\nabla\times\mathbf{A}\cdot \mathrm{d}\boldsymbol{\sigma} = q\int\mathbf{B}\cdot\mathrm{d}\boldsymbol{\sigma} = 0
\end{equation}
which is zero because the bounce motion goes along the same path in both parts of the cycle so that the enclosed area vanishes. Then, the second adiabatic invariant can be written as
\begin{equation}
 J_2 = 2\int_{s_{\rm m1}}^{s_{\rm m2}} \gamma m v_{||} \mathrm{d}s,
\end{equation}
where $s$ is the path length along the field line, and $s_{\rm m1}$, $s_{\rm m2}$ are locations of the mirror points where the particle comes back.

At the mirror point the parallel speed $v_{||}$ vanishes so that $v_\perp = v$. From the invariance of the magnetic moment $\mu$ it follows that
\begin{equation}
 \mu = \frac{\gamma^2 m v_\perp^2}{2B(s)} = \frac{\gamma^2 m v^2}{2B_{\rm m}},
\end{equation}
where $B_{\rm m}$ is the field strength at the mirror point. Substituting $v_\perp^2 = v^2-v_{||}^2$ and solving for $v_{||}$ gives
\begin{equation}
\label{eq:secinv}
J_2 = 2 \int_{s_{m1}}^{s_{m2}} \gamma m v \sqrt{1-\frac{B(s)}{B_{\rm m}}} \mathrm{d}s \equiv 2\gamma m v I.
\end{equation}
The integral $J_2$ is an adiabatic invariant in general (even if there are electric fields or slow time-dependent fields). If the speed is constant, $I$ can be used as an adiabatic invariant. The integral $I$ depends only on the magnetic field, not on the particle velocity, so it can be used to compute the drift path using the field geometry only (see exercises~\ref{ex:evaluateI}-\ref{ex:driftpath} in Section~\ref{sec:exercises}).

\begin {figure}
 \includegraphics{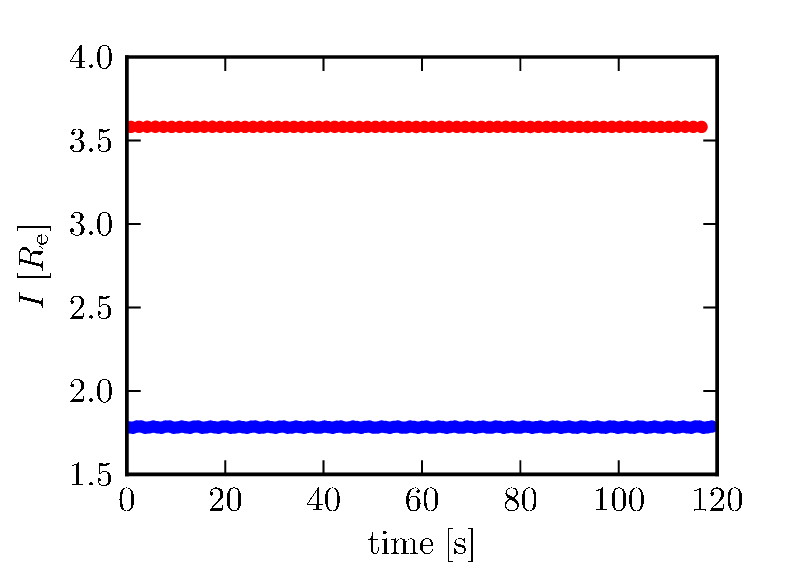}
 \caption{\label{fig:secinv}The second invariant values in time, calculated using the guiding-center trajectories in Fig.~\ref{fig:proton_gc}. Lower curve is for the proton starting at 2 $R_{\rm e}$ distance, upper curve for the proton starting at 4 $R_{\rm e}$.}
\end {figure}

Figure~\ref{fig:secinv} shows that the value of the second invariant $I$, evaluated using the guiding-center trajectories shown in Figure~\ref{fig:proton_gc}, stays constant in time. The integral is evaluated not using the definition of $I$ in Eq.~(\ref{eq:secinv}), but using the dynamical form
\begin{equation}
 I = \frac{1}{v}\int v_{||}\mathrm{d}s = \frac{1}{v}\int v_{||}^2 \mathrm{d}t,
\end{equation}
where the integral is evaluated over a half period. The limits of the integrals are determined by interpolation between two points where the parallel speed changes sign. The values do not oscillate because the adiabatic invariant is calculated as an average over a cycle.

The drift path is found by averaging the bounce motion. In a dipolar field all drift paths are circular due to the symmetry of the field. The third invariant, associated with the drift motion, is defined as an integral along the drift path:
\begin{equation}
\label{eq:thirdinv}
 J_3 = \oint (\gamma m \mathbf{v} + q\mathbf{A})\cdot \mathrm{d}\boldsymbol{\ell},
\end{equation}
where $\mathrm{d}\boldsymbol{\ell}$ is a line element on the drift path.

This can be written as
\begin{equation}
\label{eq:thirdinv2}
 J_3 = \oint \gamma m v_\mathrm{d} \mathrm{d}\mathrm{\ell} + q\int \mathbf{B}\cdot\mathrm{d}\boldsymbol{\sigma}.
\end{equation}
In the first term $v_\mathrm{d}$, the drift speed, is the magnitude of the expression given in Eq.~(\ref{eq:driftvelocity}). The second term is obtained by using Stokes' theorem as above.

An order of-magnitude comparison shows that the first term of $J_3$ can be neglected because it is much smaller than the second term: From Eq.~(\ref{eq:driftvelocity}) the order of magnitude of the drift speed can be written as
\begin{equation}
 v_{\rm d} \sim \frac{mv^2}{2qB^2}\frac{B}{R},
\end{equation}
where $B$ is the typical field strength at the drift path and $R$ is the typical distance from the origin. Similarly, from Eq.~(\ref{eq:cycrad}), the cyclotron radius has the order of magnitude
\begin{equation}
 \rho \sim \frac{mv}{qB}.
\end{equation}
Then, the order-of-magnitude ratio of the terms in Eq.~(\ref{eq:thirdinv2}) is
\begin{equation}
 \frac{mv_{\rm d}2\pi R}{qB\pi R^2} \sim \frac{m^2v^2}{q^2B^2R^2} \sim \left(\frac{\rho}{R}\right)^2.
\end{equation}
According to the adiabaticity condition Eq.~(\ref{eq:adbcondition}), $\rho/R$ must be very small. Therefore the first term of Eq.~(\ref{eq:thirdinv2}) is ignored and we have
\begin{equation}
J_3 = q\Phi,
\end{equation}
where $\Phi$ is the magnetic flux through the drift path. The third adiabatic invariant is useful as a conservation law when the magnetosphere changes slowly, i.e., over longer time scales compared to the drift period.

The use of three invariants gives more accurate results for the motion of particles over long periods. Numerical solution of equations of motion are less accurate because of accumulated numerical errors. Roederer\cite{Roederer1970} discusses in detail how drift shells can be constructed geometrically using the invariants (see Exercise~\ref{ex:driftpath} in Sec.~\ref{sec:exercises}). Furthermore, as three invariants uniquely specify a drift shell, the invariants themselves can be used as dynamical variables when investigating the diffusion of trapped particles.\cite{SchulzLanz1974}

\section{Exercises and assignments}
\label{sec:exercises}
This section lists some further programming exercises with varying difficulty. The code given in the supplement\cite{supp} can be modified to solve some of the exercises.

\begin{enumerate}

\item \textbf{Uniform magnetic field.} Follow charged particles under a uniform magnetic field $\mathbf{B}=B\hat{\mathbf{z}}$ where $B=1\mathrm{T}$. Verify that the particles follow helices with cyclotron radius and frequency as given in Eqs. (\ref{eq:cycfreq}, \ref{eq:cycrad}). Experiment with particles with different mass and charge values.

\item \label{ex:graddrift} \textbf{Gradient drift.} Consider a magnetic field given as $\mathbf{B} = (Ax+B_0)\hat{\mathbf{z}}$. The field has a gradient in the $x$-direction, but no curvature. Set $A = 1\mathrm{T\cdot m}^{-1}$ and $B_0=1\mathrm{T}$. Follow the trajectory of a particle with mass $m=1\mathrm{kg}$ and $q=1\mathrm{C}$ initialized with velocity $\mathbf{v} = 1{\rm m\cdot s}^{-1} \hat{\mathbf{y}}$ at the origin. Note that the sideways drift arises from the fact that the cyclotron radius is smaller at stronger fields.

\item \textbf{Equatorial particles.} Consider a particle in a dipolar magnetic field, located at the equatorial plane ($z=0$) with zero parallel speed. As the field strength is minimum at the equator with respect to the field line, there is no parallel acceleration and the particle stays on the equatorial plane at all times. Using the dipole model, follow an equatorial particle and verify that the center of the motion stays on a contour of constant $B$, as implied by the conservation of the first adiabatic invariant.

\item \textbf{Explore the drift motion.} Run the programs in the supplement to trace protons and electrons using the dipole model. Initialize particles with different energies, starting positions and pitch angles. Verify that electrons and protons drift in opposite directions, and electrons have a much smaller cyclotron radius than protons with the same kinetic energy. Estimate the periods of bounce and drift motions and compare them with Eqs.~(\ref{eq:bounceperiod}, \ref{eq:driftperiod}).

\item \label{ex:gcandfull} \textbf{Accuracy of the guiding-center approximation.} Simulate the full particle and guiding center trajectories with the same initial conditions and plot them together. Shift the initial position of the particle properly so that the guiding center runs through the middle of the helix.

Repeat with protons with 1keV, 10keV, 100keV and 1MeV kinetic energies. At higher energies, the guiding-center trajectory lags behind the full particle because the omitted high-order terms become more significant as the cyclotron radius increases.

\item \textbf{Different numerical methods.} Solve the full particle and guiding center equations using different numerical schemes,\cite{Garcia2000,PressEtAl2007} such as Verlet, Euler-Cromer, Runge-Kutta and Bulirsch-Stoer. Verify the accuracy of the solution by checking the conservation of kinetic energy and adiabatic invariants.

\item \textbf{Field line tracing.} Plot the magnetic dipole field line starting at position $(x_0,y_0,z_0)$. For any vector field $\mathbf{u}(\mathbf{r})$, a field line can be traced by solving the differential equation
\begin{equation}
 \frac{{\rm d}\mathbf{r}}{{\rm d}s} = \frac{\mathbf{u}}{|\mathbf{u}|}
\end{equation}
where $s$ is the arclength along the field line.

\item \label{ex:evaluateI} \textbf{Compute $I$.} Compute the second invariant $I$ (Eq.~\ref{eq:secinv}) under a dipolar field for a guiding center starting at position $(x_0,y_0,0)$ and an equatorial pitch angle $\alpha_{\rm eq}$. The integral should be taken along a field line, which can be traced as described above. From the first adiabatic invariant one finds $B_{\rm m} = B(x_0,y_0,0)/\sin^2(\alpha_{\rm eq})$ and the limits of the integral are found by solving $B(s_{\rm m}) = B_{\rm m}$. 

\item \label{ex:Ialongdrift} \textbf{Second invariant along the drift path.} Produce a guiding-center trajectory under the dipole field. By interpolation, determine the points $(x_i,y_i,0)$ where the trajectory crosses the $z=0$ plane. Compute the second invariant $I$ at these equatorial points and plot. Verify that the values are constant as shown in Fig.~\ref{fig:secinv}.

\item \label{ex:driftpath} \textbf{Drift path tracing using the second invariant.} Pick a starting location $(x_0, y_0, 0)$ and mirror field $B_{\rm m}$, and evaluate the second invariant $I(x_0,y_0,B_ {\rm m})$ as described above. Compute the gradient $\boldsymbol{\nabla}I = \partial_x I \hat{\mathbf{x}}+ \partial_y I \hat{\mathbf{y}}$ numerically using central differences:
\begin{eqnarray}
\partial_x I &\approx& \frac{1}{2\delta}\left[I(x_0+\delta,y_0,B_ {\rm m}) - I(x_0-\delta,y_0,B_ {\rm m})\right]\\
\partial_y I &\approx& \frac{1}{2\delta}\left[I(x_0,y_0+\delta,B_ {\rm m}) - I(x_0,y_0-\delta,B_ {\rm m})\right],
\end{eqnarray}
where $\delta$ is a small number (e.g. 0.01$R_{\rm e}$).

The second invariant is constant along the drift path, so for a finite step $(\Delta x,\Delta y)$, it holds that $\partial_x I \Delta x + \partial_y I \Delta y = 0$. Use this relation to trace successive steps along the drift shell. This method is more accurate than following a particle or a guiding center.

\item \textbf{The double-dipole model.}\cite{OzturkWolf2007} The dipole ceases to be a good approximation for the magnetic field of the Earth as we go farther in space. The double-dipole model, although unrealistic, introduces a day-night asymmetry that vaguely mimics the deformation of the magnetosphere by the solar wind. It can be used to capture some basic features of particle dynamics in the outer magnetosphere, if only qualitatively.

The model has one dipole (Earth) at the origin, pointing in the negative $z$-direction, and an image dipole at $x=20R_e$. If both dipoles are identical, the magnetic field is given by:
\begin{equation}
\label{eq:ddfield}
\mathbf{B}(x,y,z) = \mathbf{B}_{\rm dip}(x,y,z) + \mathbf{B}_{\rm dip}(x-20R_{\rm e},y,z) 
\end{equation}
where $\mathbf{B}_{\rm dip}(x,y,z)$ is given by Eq.~(\ref{eq:dipolefield}).

The domains of each dipole are separated by the plane $x=10R_{\rm e}$. This plane simulates the magnetopause, the boundary between the magnetosphere and the solar wind. For slightly better realism, the image dipole can be multiplied by a factor larger than 1 so that the magnetopause becomes curved. Also, the two dipoles can be tilted by equal and opposite angles with respect to the Sun-Earth line ($x$-axis), to simulate the fact that the dipole moment of the Earth is tilted.

\begin{enumerate}
\item Starting at various latitudes, plot the magnetic field lines of Eq.~(\ref{eq:ddfield}) on the $x-z$ plane. Observe the compression of field lines on the dayside and extension on the nightside. Note that no field line crosses the $x=10R_{\rm e}$ plane. Multiply the image dipole term by 1.5 and repeat.

\item Follow several guiding-center trajectories starting position $x$ between $-7R_{\rm e}$ and $-10R_{\rm e}$, $y=z=0$, and pitch angles between $30^\circ$ and $60^\circ$ (smaller pitch angle creates a longer bounce motion). With small pitch angles, the particle should come closer to Earth on the day side. Repeat with a pitch angle of $90^\circ$. Now the particle goes away from the Earth on the dayside. Explain these observations using the conservation of first and second adiabatic invariants.

\item The double-dipole field can break the second invariant for some trajectories. The reason of this breaking is that the field strength has a local maximum on the dayside around the equatorial plane. Particles with sufficiently small mirror fields are diverted to one side of the equatorial plane because they cannot overcome this field maximum, as seen in Fig.~\ref{fig:protonDSB}.

\begin{figure}
\includegraphics{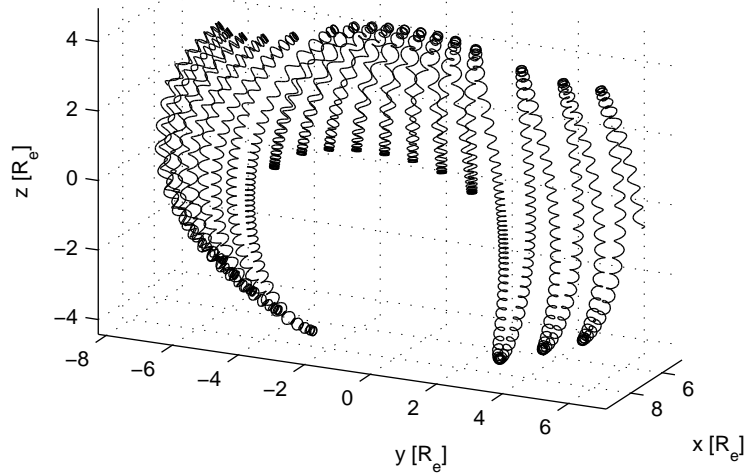}
\caption{\label{fig:protonDSB}A proton with 200keV kinetic energy, initialized on the right edge at $(7R_{\rm e}, 7R_{\rm e}, 0)$ with $60^\circ$ pitch angle in a double-dipole field.}
\end{figure}

Start an electron guiding center at position $x_0 = -10R_{\rm e}$, $y_0=z_0=0$ with kinetic energy $K=1\mathrm{MeV}$ with an equatorial pitch angle $70^\circ$ and follow its guiding center for 1000 seconds with time step 0.01. On the dayside the trajectory will temporarily move above or below the equatorial plane. Using the method used in Fig.~\ref{fig:secinv}, plot the second invariant $I$ versus time. The second invariant will be constant between breaking points, but its value will differ from the initial value. The reason is that near the breaking points bounce motion slows down and the adiabaticity condition does not hold. However, the first invariant is not broken.

This phenomenon, named drift-shell bifurcation,\cite{OzturkWolf2007} can be one of the causes of particle diffusion in the magnetosphere.
\end{enumerate}

\item \textbf{Magnetotail current sheet.} On the tail region of the magnetosphere, magnetic field lines are heavily stretched, and a sheet of current is flowing through them.\cite{Hughes1995} The field in the magnetotail can be represented by the simple form:
\begin{eqnarray}
\mathbf{B} &=& B_0\frac{z}{d}\hat{\mathbf{x}} + B_n \hat{\mathbf{z}} \text{\ \ if\ \ }|z|<1 \nonumber\\
&=& \frac{B_0}{d}\hat{\mathbf{x}} + B_n \hat{\mathbf{z}} \text{\ \ otherwise}
\end{eqnarray}
In this problem we set $B_0=10$, $B_n=1$ and $d=0.2$. The field lines trace parabolas on the $x-z$ plane, which can be seen by integrating the equation $\mathrm{d}x/B_x = \mathrm{d}z/B_z$. The parameter $d$ is the scale of the current sheet thickness. The truncation of the field at $z=1$ simulates the finite size of the tail region.

The field vector points to opposite directions on both sides of the equatorial plane $z=0$. When a charged particle is released from above, it moves toward the weaker region near $|z|=0$ where the adiabaticity condition does not hold. The helix becomes a ``serpentine orbit'' that moves in and out of the equatorial plane. The chaotic dynamics of these orbits is extensively studied.\cite{Chen1992,BuchnerZelenyi1989}

\begin{figure}
\includegraphics{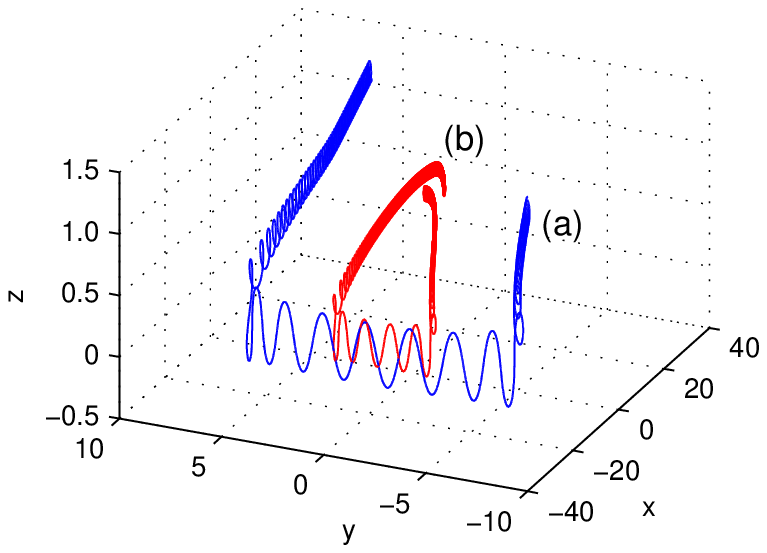}
\includegraphics{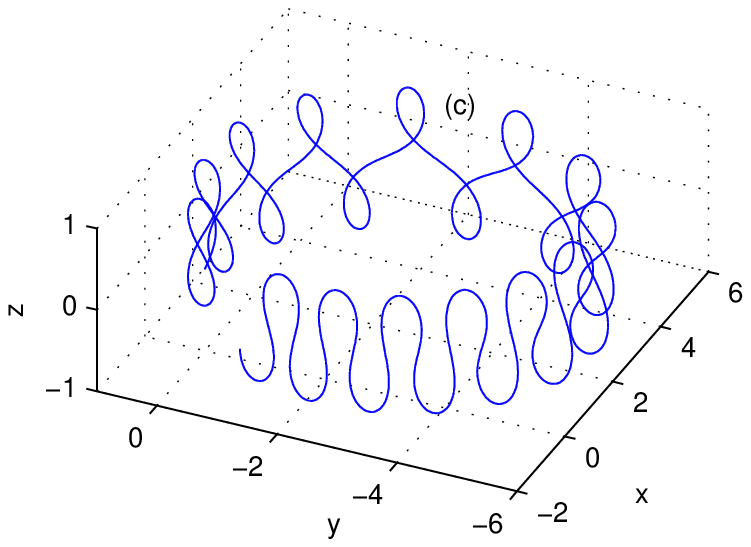}
 \caption{\label{fig:Speiser}(Color online) Types of orbits created by a particle with mass $m=5$ and charge $q=1$ near a current sheet. (a) Speiser orbits of transient particles, (b) Cucumber orbits of quasitrapped particles and (c) Ring orbits of trapped particles. Note the different scales of axes.} 
\end{figure}

Figure~\ref{fig:Speiser} shows the three types of orbits that can exist in such a model.\cite{BuchnerZelenyi1989,SharmaEtAl2006} ``Speiser orbits'' approach the equatorial plane and later go beyond $z=1$ and leave the tail region. ``Cucumber orbits'' alternate between helical and serpentine orbits. These do not form closed orbits because of the breaking of the first invariant at the equatorial plane. ``Ring orbits'' alternate between oppositely-directed fields; they do not have a helical section.

\begin{enumerate}

\item Evaluate $\boldsymbol{\nabla}B$ for $|z|<1$ and determine the direction of gradient-curvature drift.

\item By trial and error, find initial conditions that create the types of orbits shown in Fig.~\ref{fig:Speiser}.
\end{enumerate}

\end{enumerate}

\section{Concluding remarks}
\label{sec:conclusion}
Space plasmas provide many case studies which, after proper simplification, can be used in the undergraduate physics curriculum. We have presented one such case, the basic theory of charged-particle motion under the dipole.

This paper focuses on visualization and concrete computation, with the hope that students will modify or rewrite the code to run their own numerical experiments on particle motion in magnetic fields. In my opinion, numerical simulations provide at least two important pedagogical benefits: First, even if the required analytical tools are beyond the students' level, they can use simulations to obtain a qualitative understanding. Second, the process of coding the simulation forces students to understand the problem at a basic and operational level.

The main body of this article or the exercises can be incorporated in lectures, or they can be given as advanced assignments to interested students. A natural place for this subject is a course on electromagnetism and/or plasma physics. When the basics are introduced, the instructor can discuss related subjects such as plasma confinement, radiation belts, or space weather.

In advanced mechanics courses, adiabatic invariants are usually presented with an abstract formalism. Charged particle motion provides a natural and concrete case where adiabatic invariants are relevant and indispensable.

The subject can also be incorporated in courses on computational physics. Accuracy and stability of different numerical integration schemes may be presented using charged particle motion. The widely separated time scales of the motion would be a challenge for most of the schemes.

\begin{acknowledgments}
I thank Meral \"Ozt\"urk for her careful proofreading, and two anonymous reviewers for their helpful suggestions.
\end{acknowledgments}


\begin{thebibliography}{99}
\bibitem{note:neglectofE} Electric potential differences across the magnetosphere are of the order of 100kV which, over a distance of about $20R_{\rm e}$, yield an electric field strength of the order of $E \sim 10^{-3}{\rm Vm}^{-1}$. The magnetic field of the Earth has strength $3.07\times10^{-5}\mathrm{T}/L^3$ on the magnetic equatorial plane, where $L$ is the distance measured in $R_{\rm e}$. A typical radiation-belt proton with 10MeV energy has speed $v \sim 0.1c$. Then, in SI units, $vB\sim 1000/L^3\gg E$. Therefore the electric drift can be neglected near the Earth where $L<6$. Alternatively one can say that low-energy particles that could be affected by electric drifts have already drifted away, leaving behind the trapped high-energy particles.
\bibitem{note:GSE} We are using the Geocentric Solar Ecliptic (GSE) coordinate system. The Earth is at the origin, the $x$-axis points to the Sun, the $z$-axis is set so that the dipole moment vector is on the $xz$-plane, and the $y$-axis is perpendicular to both axes.
\end{thebibliography}
 \end{document}